\documentclass[aps,pra,twocolumn,superscriptaddress]{revtex4-2}
\usepackage[utf8]{inputenc}
\usepackage{graphicx}
\usepackage{amsmath}
\usepackage{natbib}
\usepackage{subcaption}
\usepackage{float}
\usepackage{xcolor}
\usepackage{placeins}
\usepackage{braket}
\begin{document}

\title{Ergotropy Dynamics in a Dissipative Graphene Quantum Battery}
\author{Disha Verma}
\email{vermadisha785@gmail.com}
\affiliation{Department of Physics, National Institute of Technology, Tiruchirappalli 620015, India.}

\author{Indrajith VS}
\affiliation{Qdit Labs Pvt. Ltd., Amruthahalli, Bangalore 560092, India.}

\author{R. Sankaranarayanan}
\affiliation{Department of Physics, National Institute of Technology, Tiruchirappalli 620015, India.}

\begin{abstract}
We investigate ergotropy dynamics in a graphene-based quantum battery modeled as a four-level spin--valley system under different dissipative environments. The battery is charged via a Gaussian pulse and subsequently evolves under amplitude damping, dephasing, and both Markovian and non-Markovian reservoirs. We find that amplitude damping, while inducing energy loss, can stabilize non-passive steady states with finite ergotropy, whereas pure dephasing suppresses coherence and eliminates work extraction. On the other hand, non-Markovian memory slows ergotropy loss and enables partial recovery through information backflow. These results identify coherence and reservoir memory as essential resources for enhancing the long-time performance of graphene quantum batteries.
\end{abstract}

\maketitle

\section{Introduction}
Rapid advancements in quantum information processing have fueled the search for novel nanomaterials that can operate within the quantum regime. Among these, graphene stands out as a promising platform due to its exceptional quantum properties, including high carrier mobility and long decoherence times, which make it ideal for quantum technologies. Understanding and exploiting quantum resources in graphene-based systems has therefore become a topic of growing interest.

Graphene-based platforms and related nanoscale systems have been widely explored in diverse contexts, including sensing, photonic, and energy applications \cite{lu2024peptide, peng2025ultrahigh, zhang2025dominant}, highlighting their versatility for emerging quantum and energy-storage technologies.Graphene is a two-dimensional material composed of carbon atoms arranged in a honeycomb lattice, first isolated by Geim and Novoselov in 2004 \cite{novoselov2004electric}. Its unique electronic properties arise from massless Dirac fermions originating from the two-sublattice structure of the lattice \cite{semenoff1984condensed, divincenzo1984self, fradkin1986critical, haldane1988model}. The low-energy states are concentrated near the inequivalent \(K\) and \(K'\) points in the Brillouin zone, introducing a valley degree of freedom. This valley pseudospin, combined with real spin, results in a rich internal state structure that supports exotic quantum phenomena such as the anomalous quantum Hall effect and minimal conductivity \cite{novoselov2004electric, novoselov2005two}. The low-energy behavior of carriers in graphene is well described by a Dirac-like Hamiltonian derived from tight-binding approximations.

Quantum batteries (QBs) have emerged as a novel concept in quantum thermodynamics, capable of storing and delivering energy by harnessing quantum coherence and correlations. Theoretical studies have proposed a variety of QB models, including spin chains, cavity-QED systems, Gaussian and free-fermion batteries, dissipative charging protocols, and many-body interacting setups \cite{verma2025dynamics, le2018spin, dou2022cavity, gemme2023off, catalano2024frustrating, cavaliere2025gaussian, grazi2025fermion, massa2025collisional, cavaliere2025blockade, grazi2026quenches, chand2026spin, farina2026phase, pavone2026cluster, grazi2025finite}. The role of dissipation and environmental noise in quantum battery performance has also been actively investigated in recent studies \cite{HU2025116229,kamin2020nonmarkovian,zhao2021environmental,zakavati2021bounds,quach2020dark}. Similar to classical batteries, a QB operates through three main stages: charging, storage, and discharging. Charging is typically achieved via a time-dependent interaction Hamiltonian, \( \mathcal{H}_1 \), applied over a finite duration \( \tau_c \), during which the system \( \mathcal{H}_0 \) evolves according to the time-dependent Schrödinger equation. When \( [\mathcal{H}_0, \mathcal{H}_1] \neq 0 \), the charging dynamics are non-trivial, enabling energy transfer into the system.

In realistic scenarios, QBs are open quantum systems due to unavoidable interactions with their environments.To address this, various studies have explored robust states and protocols for mitigating decoherence, dissipative stabilization, noisy charging dynamics, and retaining extractable work (ergotropy) \cite{quach2020using, liu2019loss}. A detailed review of open-system QBs can be found in \cite{campaioli2024colloquium}.

To ensure the physical relevance of our model, we emphasize that the spin–valley tensor states in graphene are not mere theoretical constructs but have been realized and controlled experimentally.Graphene nanostructures, including quantum dots and SNS (superconductor-normal metal-superconductor) junctions, have demonstrated tunable spin and valley degrees of freedom \cite{eich2018spin, yang2013spin}. Moreover, proximity-induced superconductivity in graphene SNS junctions enables phase-sensitive manipulation of Andreev bound states carrying spin and valley indices \cite{beenakker2006specular, traversoziani2015josephson, cavaliere2018andreev}. Recent studies on hybrid superconducting nanostructures and mesoscopic quantum transport have further highlighted the role of Andreev physics and coherent transport mechanisms in graphene-based quantum devices. Valleytronics studies have further shown that the \(K/K'\) valleys can serve as computational bases, with selective excitation achieved using circularly polarized light \cite{mak2014valley}. These developments provide a strong experimental foundation for graphene-based quantum battery designs.

Motivated by these advances, we propose a four-level graphene quantum battery model based on spin–valley coupled qubits, where charging is induced via a Gaussian pulse in a closed system. Following the charging phase, we investigate the effects of environmental dissipation modeled by both Markovian and non-Markovian amplitude damping channels. This approach allows us to evaluate the intrinsic performance of the battery, as well as the retention of ergotropy under realistic open-system conditions. Similar pulse-based charging protocols and dissipation analyses have been studied in the context of quantum thermodynamics \cite{binder2018thermodynamics, paulson2025work, barra2019dissipative}, providing a firm theoretical foundation for our work.

\section{Model}
\subsection{Graphene-Based Quantum Battery}\label{the}

In monolayer graphene, the low-energy electronic structure arises from two inequivalent sublattices (A and B) and two valleys (K and K$'$) in the Brillouin zone. These internal degrees of freedom can be described using two sets of Pauli matrices: 
$\vec{\sigma}=(\sigma_x,\sigma_y,\sigma_z)$ acting on the sublattice pseudospin, and 
$\vec{\tau}=(\tau_x,\tau_y,\tau_z)$ acting on the valley pseudospin. 
The combined Hilbert space is four-dimensional, spanned by the basis 
$\{\ket{00},\ket{01},\ket{10},\ket{11}\}$, 
where the first (second) index labels the sublattice (valley) degree of freedom.  Within the effective-mass approximation near the Dirac points, the low-energy Hamiltonian is written as~\cite{hu2009quantum}
\begin{equation}
H_0 = \eta \!\left[n_x(\sigma_x \!\otimes\! I_2) + n_y(\sigma_y \!\otimes\! \tau_z)\right] 
+ H_{\text{int}},
\label{eq:k1}
\end{equation}
where $\eta$ sets the kinetic energy scale related to the Fermi velocity, and  $n_x,n_y$ introduce momentum-dependent asymmetry between the off-diagonal couplings. $H_{\text{int}}$ incorporates inter-sublattice and inter-valley couplings.

The interaction between the sublattice and valley pseudospins in graphene can be expressed as
\begin{equation}
H_{\text{int}} = 
\lambda \left( 
I_4 
+ e^{-i\alpha} \sigma_{+} \otimes I_2 
+ e^{i\alpha} \sigma_{-} \otimes I_2
\right),
\label{eq:Hint_eq}
\end{equation}
where $\lambda$ sets the characteristic interaction strength and $\alpha$ represents an anisotropy phase between inter-sublattice couplings. Here, $\sigma_{\pm} = (\sigma_x \pm i\sigma_y)/2$ act on the sublattice pseudospin space, while $I_2$ is the identity in the valley subspace. The first term, proportional to $I_4$, corresponds to on-site intravalley contributions, whereas the remaining terms describe inter-sublattice tunneling processes with a complex phase factor $e^{\pm i\alpha}$. This phase controls the direction-dependent hybridization between sublattice states, introducing an effective anisotropy in the coupling. 
Combining Eqs.~\eqref{eq:k1}--\eqref{eq:Hint_eq} gives the effective four-level Hamiltonian matrix
\begin{equation}
H_0 = 
\lambda \,
\scalebox{0.85}{$
\begin{pmatrix}
1 & e^{-i\alpha} & \tilde{\eta}(n_x - i n_y) & 0 \\
e^{i\alpha} & 1 & 0 & \tilde{\eta}(n_x + i n_y) \\
\tilde{\eta}(n_x + i n_y) & 0 & 1 & e^{-i\alpha} \\
0 & \tilde{\eta}(n_x - i n_y) & e^{i\alpha} & 1
\end{pmatrix}
$},
\label{eq:H1}
\end{equation}
where $\tilde{\eta}=\eta/\lambda$ quantifies the relative coupling strength.  

Diagonalization of Eq.~\eqref{eq:H1} yields four eigenstates :
\begin{equation}
\begin{aligned}
\ket{\phi_1} &= e^{-i\alpha} A_{+} \ket{00}
+ \tfrac{1}{2} \ket{01}
+ \tfrac{1}{2} e^{-i\alpha} \ket{10}
+ A_{+} \ket{11}, \\
\ket{\phi_2} &= -e^{-i\alpha} A_{+} \ket{00}
+ \tfrac{1}{2} \ket{01}
+ \tfrac{1}{2} e^{-i\alpha} \ket{10}
- A_{+} \ket{11}, \\
\ket{\phi_3} &= -\tfrac{1}{2} e^{-i\alpha} \ket{00}
+ A_{-} \ket{01}
- e^{-i\alpha} A_{-} \ket{10}
+ \tfrac{1}{2} \ket{11}, \\
\ket{\phi_4} &= -\tfrac{1}{2} e^{-i\alpha} \ket{00}
- A_{-} \ket{01}
+ e^{-i\alpha} A_{-} \ket{10}
+ \tfrac{1}{2} \ket{11}.
\end{aligned}
\label{eq}
\end{equation}
where
\begin{equation}
A_{\pm} = -\frac{k_{1} \pm 1 \mp i k_{2}}{2 \sqrt{1 + k_{1}^2 \pm 2k_{1} + k_{2}^2}}, 
\qquad
k_{1} = \frac{\eta n_x}{\lambda}, \;
k_{2} = \frac{\eta n_y}{\lambda}.
\end{equation}
The corresponding energy eigenvalues are
\begin{align}
E_1 &= \lambda \!\left(1 - \sqrt{1 + k_{1}^2 + 2k_{1} + k_{2}^2}\right), \\
E_2 &= \lambda \!\left(1 + \sqrt{1 + k_{1}^2 + 2k_{1} + k_{2}^2}\right), \\
E_3 &= \lambda \!\left(1 - \sqrt{1 + k_{1}^2 - 2k_{1} + k_{2}^2}\right), \\
E_4 &= \lambda \!\left(1 + \sqrt{1 + k_{1}^2 - 2k_{1} + k_{2}^2}\right).
\end{align}

These four eigenvalues define the discrete energy spectrum of the graphene-based quantum battery. Throughout this analysis, $\eta$, $\lambda$, $n_x$, and $n_y$ are treated as dimensionless quantities for convenience.

In experimentally realizable graphene-based platforms, such as gate-defined quantum dots and proximitized graphene heterostructures, these degrees of freedom can be selectively controlled via electrostatic gating, magnetic fields, or proximity-induced spin--orbit coupling \cite{eich2018spin,beenakker2006specular} . The parameters $\lambda$, $\eta$, and $\alpha$ in Eq.~(\ref{eq:H1}) effectively capture these tunable interactions, including anisotropic coupling and phase-dependent hybridization between sublattice states. Within this framework, the four-level system acts as a minimal quantum battery, where energy is stored in the occupation and coherence of these spin--valley states and can be manipulated via external driving fields, such as the Gaussian pulse introduced in Eq.~(\ref{eq:Hpulse}).
This mapping establishes that the present model is not purely abstract, but represents a physically motivated and experimentally relevant description of energy storage in graphene-based quantum systems.

\subsection{Charging Protocol}

The battery is initialized in the ground state of the system Hamiltonian $H_0$, obtained by solving:
\begin{equation}
    H_0|\psi_{GS}\rangle = E_{GS}|\psi_{GS}\rangle,
\end{equation}
with the corresponding density matrix given by $\rho_0 = |\psi_{GS}\rangle\langle\psi_{GS}|$.

The charging process is driven by an external, time-dependent Gaussian pulse field represented as
\begin{equation}
H_{\text{pulse}}(t) = B_s
\begin{pmatrix}
1 & 0 & 0 & 0 \\
0 & -1 & 0 & 0 \\
0 & 0 & 1 & 0 \\
0 & 0 & 0 & -1
\end{pmatrix}
e^{-t^2/(2\tau^2)},
\label{eq:Hpulse}
\end{equation}
The chosen pulse Hamiltonian is analogous to a longitudinal Zeeman interaction, where the diagonal form induces opposite energy shifts for states of opposite spin or pseudospin orientation. Such a coupling alters the energy spectrum without causing inter-state transitions, thereby preserving phase coherence during the charging dynamics. This results in a coherent, reversible energy injection process that can be experimentally realized through spin- or valley-selective electrostatic or magnetic field in graphene-based systems. Here, $t$ denotes time, governing the temporal evolution of the charging process, while $\tau$ characterizes the pulse width and thus the duration of the external drive. 

A smaller $\tau$ corresponds to a faster, more localized pulse. In this study, $\tau = 1.0$ is chosen as a normalized time scale to represent a fast charging process within a short interaction window. A larger $B_s$ produces stronger driving and faster population transfer between energy levels.

Gaussian pulses are commonly used in quantum control and thermodynamic processes because they are experimentally feasible and minimize abrupt transitions that could cause unwanted excitations~\cite{quach2022superabsorption, ferraro2018high, campaioli2017enhancing}. The total time-dependent Hamiltonian governing the charging process is thus
\begin{equation}
H(t) = H_0 + H_{\text{pulse}}(t),
\label{eq:Htotal}
\end{equation}.  
The applied pulse coherently drives transitions between eigenstates of $H_0$, effectively transferring energy into the system and charging the quantum battery. We note that the Gaussian pulse is chosen for its smooth temporal profile and experimental feasibility,  while the Hamiltonian parameters and pulse characteristics govern the efficiency and dynamics of the charging process, the primary focus of the present work is on the dissipative evolution of the system following the charging stage. In this context, these parameters determine the initial non-passive state from which open-system dynamics proceeds. A detailed optimization of coupling strength and pulse width for maximizing charging performance is beyond the scope of the present study and is left for future investigation.

\section{Observables}

To analyze the charging and dissipation dynamics of the graphene quantum battery, we evaluate key physical quantities that characterize its energetic and quantum properties. These include the stored energy, fidelity-based purity, and the ergotropy, which together provide a comprehensive picture of the battery’s performance under different environmental interactions.

\subsection{Energy and Purity}

The stored energy of the quantum battery is defined relative to the
ground-state energy of the battery Hamiltonian $H_0$ as
\begin{equation}
\Delta E(t)
=
E{(t)}
-
E_{\mathrm{gs}},
\label{eq:stored_energy}
\end{equation}
where $E_{\mathrm{gs}}$ denotes the
ground-state energy and $E(t) = Tr[H_{0}  \rho(t)]$ denotes the energy at any arbitrary time $t$. This definition removes arbitrary energy offsets
and ensures that the stored energy quantifies the physically accessible
energy above the passive ground-state reference, consistent with the
ergotropy bound $\mathcal{W}(t) \le \Delta E(t)$, 

To quantify the purity, we employ the fidelity-based purity measure  ~\cite{indrajith2022fidelity}, defined as
\begin{equation}
    P_F(\rho) = \log_d \!\left( d \, \text{Tr}[\rho(t)^2] \right),
    \label{eq:purity}
\end{equation}
where \( d \) is the Hilbert-space dimension of the system. This purity measure indeed has intricate connection to the quantum coherence.
This logarithmic normalization renders $P_F(\rho)$ dimension-independent and bounded, providing an intuitive interpretation of mixedness: $P_F(\rho)=1$ for a perfectly coherent pure state and $P_F(\rho)<1$ as mixedness sets in.  
Together, the quantities $\langle H_0 \rangle$ and $P_F(\rho)$ capture the interplay between energetic retention and coherence degradation that governs the operational efficiency of the battery.

\subsection{Ergotropy}

The extractable work content of a quantum state is quantified by its \emph{ergotropy}, which represents the maximum amount of work obtainable through unitary operations without changing the state’s entropy.  
For a time-dependent Hamiltonian $H(t)$ and a corresponding density matrix $\rho(t)$, the ergotropy is defined as
\begin{equation}
    \mathcal{E}(t) = \text{Tr}[H(t)\rho(t)] - \text{Tr}[H(t)\rho_{\text{passive}}(t)],
    \label{eq:ergotropy}
\end{equation}
where $\rho_{\text{passive}}(t)$ is the passive counterpart of $\rho(t)$ with respect to $H(t)$. 
The passive state $\rho_{\text{passive}}(t)$ is constructed by arranging the eigenvalues of the density matrix $\rho(t)$ in decreasing order and assigning them to the eigenstates of $H(t)$ in increasing order of energy:
\begin{equation}
\rho_{\text{passive}}(t) = \sum_i p_i^{\downarrow} \ket{E_i^{\uparrow}}\bra{E_i^{\uparrow}},
\label{eq:passive_state}
\end{equation}
where $p_i^{\downarrow}$ are the eigenvalues of $\rho(t)$ sorted in descending order, and $\ket{E_i^{\uparrow}}$ are the eigenstates of $H(t)$ arranged in ascending order of energy~\cite{allahverdyan2004maximal, francica2017daemonic}.  
The passive state shares the same spectrum as $\rho(t)$ but has zero extractable work.

For a closed system, the evolution is purely unitary,
\begin{equation}
    \rho(t) = U(t)\rho_0 U^{\dagger}(t),
    \label{eq:unitary_evolution}
\end{equation}
with
\begin{equation}
    U(t) = \mathcal{T} \exp\!\left[-i \!\int_0^t \! H(t)\, ds\right],
\end{equation}
where $\mathcal{T}$ denotes time ordering.  
Since entropy remains constant under unitary evolution, any energy gain directly contributes to ergotropy.

In open-system scenarios, interactions with an external environment induce non-unitary evolution governed by the Lindblad master equation,
\begin{equation}
    \dot{\rho}(t) = -i[H(t), \rho(t)] +  \gamma \mathcal{D}[L]\rho(t),
    \label{eq:lindblad}
\end{equation}
where $\mathcal{D}[L]\rho = L\rho L^{\dagger} - \tfrac{1}{2}\{L^{\dagger}L, \rho\}$ denotes the dissipator associated with the channel $L$, $\gamma$ represents the dissipation strength and $\{A,B\} = AB + BA$ represents the anticommutator between operators $A$ and $B$. The evolution of the system’s density matrix \( \rho(t) \) is governed by the Lindblad master equation~\cite{breuer2002theory, manzano2020short, nielsen2010quantum}. The instantaneous mixed state $\rho(t)$ obtained by numerically solving eq.(\ref{eq:lindblad}) is then used to compute the ergotropy $\mathcal{E}(t)$ at each time step.

Thus, by jointly analyzing $\langle H_0 \rangle$, $P_F(\rho)$, and $\mathcal{E}(t)$, we obtain a complete dynamical characterization of the battery’s energy retention, coherence evolution, and useful work capacity under realistic dissipative conditions.

\section{Results}

The parameters of the Hamiltonian in eq.(\ref{eq:k1}) are chosen to ensure a balanced and physically realistic energy landscape for the four-level graphene quantum battery. The coupling constant $\lambda$ sets the overall energy scale, while the phase parameter $\alpha$ introduces a complex phase factor ($e^{\pm i\alpha}$) in the interaction terms of eq.(\ref{eq:H1}). This phase breaks the perfect symmetry between the two graphene sublattices, inducing controlled anisotropy in the coupling strengths. Setting $\alpha = \pi/4$ yields a moderate imbalance that preserves coherence while generating nontrivial interference between transition pathways—an essential feature for sustaining quantum coherence and nonzero ergotropy during the charging process.

The ratio $\eta = 0.5$ defines the relative strength between the kinetic and interaction contributions, ensuring a stable and well-separated eigenvalue spectrum. The asymmetric choice of momentum components, $n_x = 1$ and $n_y = 5$, represents anisotropic quasiparticle propagation within the graphene plane. This asymmetry breaks degeneracy among the energy levels, enriching the dynamical response of the system and enhancing the diversity of accessible transition channels. Collectively, these parameter choices capture realistic graphene band anisotropy and coherent coupling while maintaining numerical stability and analytical tractability.

Charging of the graphene quantum battery is achieved through a time-dependent Gaussian pulse field described by the Hamiltonian in eq.(\ref{eq:Hpulse}). At $t=5\tau$, the pulse amplitude is reduced by a factor
$e^{-25/2}\approx 3\times10^{-6}$, and at $t=10\tau$ it is
$e^{-50}\sim10^{-22}$.
Hence the pulse is effectively zero well before the final simulation time. The system is initially prepared in the ground state of $H_0$ and subsequently evolves under the total Hamiltonian $H(t)$ during the charging phase. As the process is closed and unitary, the evolution follows eq.(\ref{eq:unitary_evolution}), ensuring reversible and coherent energy storage.

\subsection{Effect of Dissipation Strength }
\label{subsec:erg}

After charging, the quantum battery interacts with its environment, leading to decoherence. In our model, the environment is described by a Markovian amplitude damping (AD) channel characterized by spontaneous emission without thermal excitation. This corresponds to a zero-temperature reservoir, where population can decay from the excited to ground state but not vice versa.  The corresponding collapse operator for the two-qubit system is defined as:
\begin{equation}
    L_{\text{AD}} =  \left( \sigma_- \otimes I + I \otimes \sigma_- \right),
    \label{eq:ad}
\end{equation}
where \( \sigma_- \) is the lowering (annihilation) operator for a single qubit. We employ a local master equation where amplitude damping
acts independently on each spin degree of freedom. This approach captures scenarios where each subsystem couples to an independent bath and the inter-spin interaction energy scale is comparable to or smaller than the system–bath coupling strength.

The influence of dissipation strength on the graphene quantum battery reveals a subtle interplay between energy retention, coherence decay, and ergotropy stabilization. To quantify coherence, we employ the $\ell_1$-norm measure \cite{baumgratz2014quantifying}, defined as
\begin{equation}
C_{\ell_1}(\rho) = \sum_{i \neq j} \big|\rho_{ij}\big|,
\end{equation}
which captures the contribution of all off-diagonal terms of the density matrix.(Refer to appendix A,B and C)

\begin{figure}[H]
    \centering
    \includegraphics[width=0.9\linewidth]{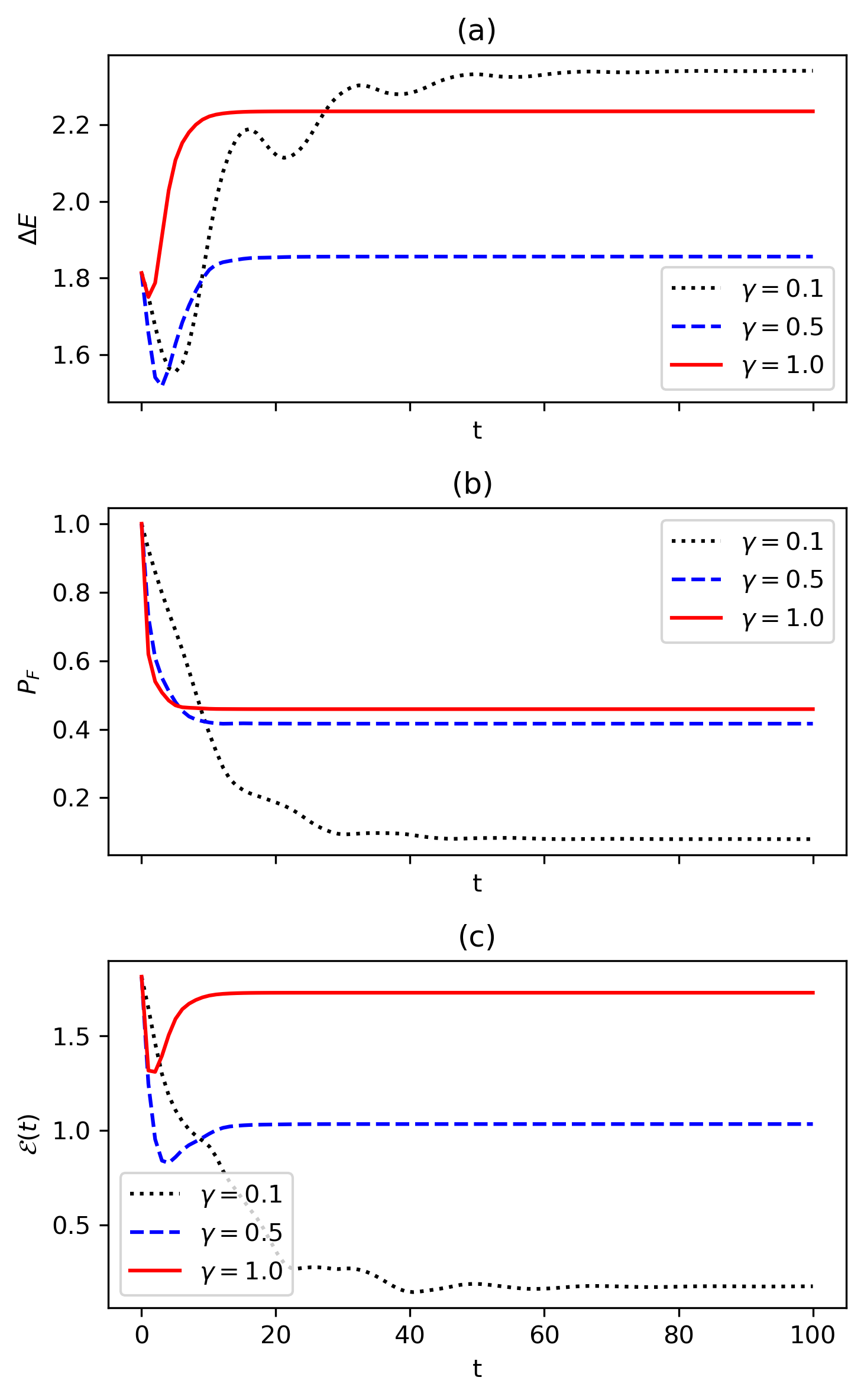}
    \caption{Time evolution of (a) energy, (b) purity, and (c) ergotropy for different dissipation strengths $\gamma = 0.1$, $0.5$, and $1.0$. The time grid comprises $1000$ uniformly spaced points over 
the interval $t \in [0,100]$, yielding an output sampling interval 
$\Delta t \approx 0.1$. The Lindblad master equation is solved 
using an adaptive-step ODE integrator. Numerical convergence 
was confirmed by decreasing the sampling interval to 
$\Delta t \approx 0.01$.}
    \label{fig:dissipation_analysis}
\end{figure}

The overall analysis of Fig.~\ref{fig:dissipation_analysis} shows that for \textbf{weak dissipation} ($\gamma = 0.1$), the stored energy builds up during charging as the dissipation strength is weak, leading to a high energy storage with time. Purity decreases gradually, reflecting persistent but slow decoherence. Inspection of the density matrices shows that the off-diagonal components (responsible for quantum coherence) diminish strongly, with $\ell_1$-coherence dropping from $C_{\ell_1} \approx 2.10$ at $t = 0$ to $C_{\ell_1} \approx 1.37$ at $t = 10$, and further down to $C_{\ell_1} \approx 0.47$ by $t = 100$. Meanwhile, diagonal populations redistribute toward near-equilibrium values, rendering the state close to passive. Consequently, the ergotropy decays from $\mathcal{E}(t) \approx 1.81$ at $t = 0$ to $\mathcal{E}(t) \approx 0.92$ at $t = 10$, and finally vanishes asymptotically ($\mathcal{E}(t) \approx 0.17$ at $t = 100$), even though residual stored energy remains.

For \textbf{intermediate dissipation} ($\gamma = 0.5$), the system loses energy more rapidly than in the weak case, but equilibration occurs at a finite steady value rather than full decay. Purity degrades faster, indicating stronger system--environment entanglement. Coherence is reduced by nearly half, from $C_{\ell_1} \approx 2.10$ at $t = 0$ to $C_{\ell_1} \approx 1.00$ by $t = 40$--$100$, while diagonal 
populations remain noticeably imbalanced. This incomplete thermalization preserves a moderate ergotropy plateau: $\mathcal{E}(t)$ drops sharply to $\approx 0.98$ at $t = 10$, but then stabilizes at $\mathcal{E}(t) \approx 1.03$ for long times ($t \ge 40$). Thus, intermediate damping achieves a balance: destructive coherences are suppressed, while population asymmetries survive long enough to sustain useful work extraction.

In the case of \textbf{strong dissipation} ($\gamma = 1.0$), the stored energy relaxes quickly but saturates at a higher steady value than in the $\gamma = 0.5$ case. This counterintuitive effect arises  from dissipation-assisted stabilization: fast damping suppresses harmful oscillations and selectively freezes favorable population distributions. Purity drops sharply at short times and then stabilizes. Although coherence collapses rapidly ($C_{\ell_1} \approx 2.10$ at $t = 0$, to $C_{\ell_1} \approx 0.89$ already by $t = 10$, and then remains nearly constant thereafter), the diagonal populations stabilize 
in a robustly non-passive arrangement, where $\rho_{00}$ and $\rho_{22}$ dominate over $\rho_{11}$ and $\rho_{33}$. This locked imbalance maintains a significant fraction of extractable work, explaining why ergotropy stabilizes at the highest asymptotic value, $\mathcal{E}(t) \approx 1.73$ (observed from $t = 40$ onward).

Overall, these results demonstrate that dissipation does not universally degrade performance. Weak dissipation prolongs coherence but ultimately drives the system to passivity and negligible ergotropy. Intermediate dissipation balances coherence suppression with partial population retention, yielding moderate work. Strong dissipation, paradoxically, enhances long-time ergotropy 
by eliminating harmful coherences while preserving favorable population imbalances. This dissipation-assisted stabilization mechanism, also observed in related studies \cite{ccakmak2020ergotropy,francica2017daemonic, ghosh2021fast}, highlights the possibility of harnessing structured environments to optimize ergotropy retention and work capacity in graphene-based quantum batteries.

\begin{figure}[H]
    \centering
    \includegraphics[width=0.9\linewidth]{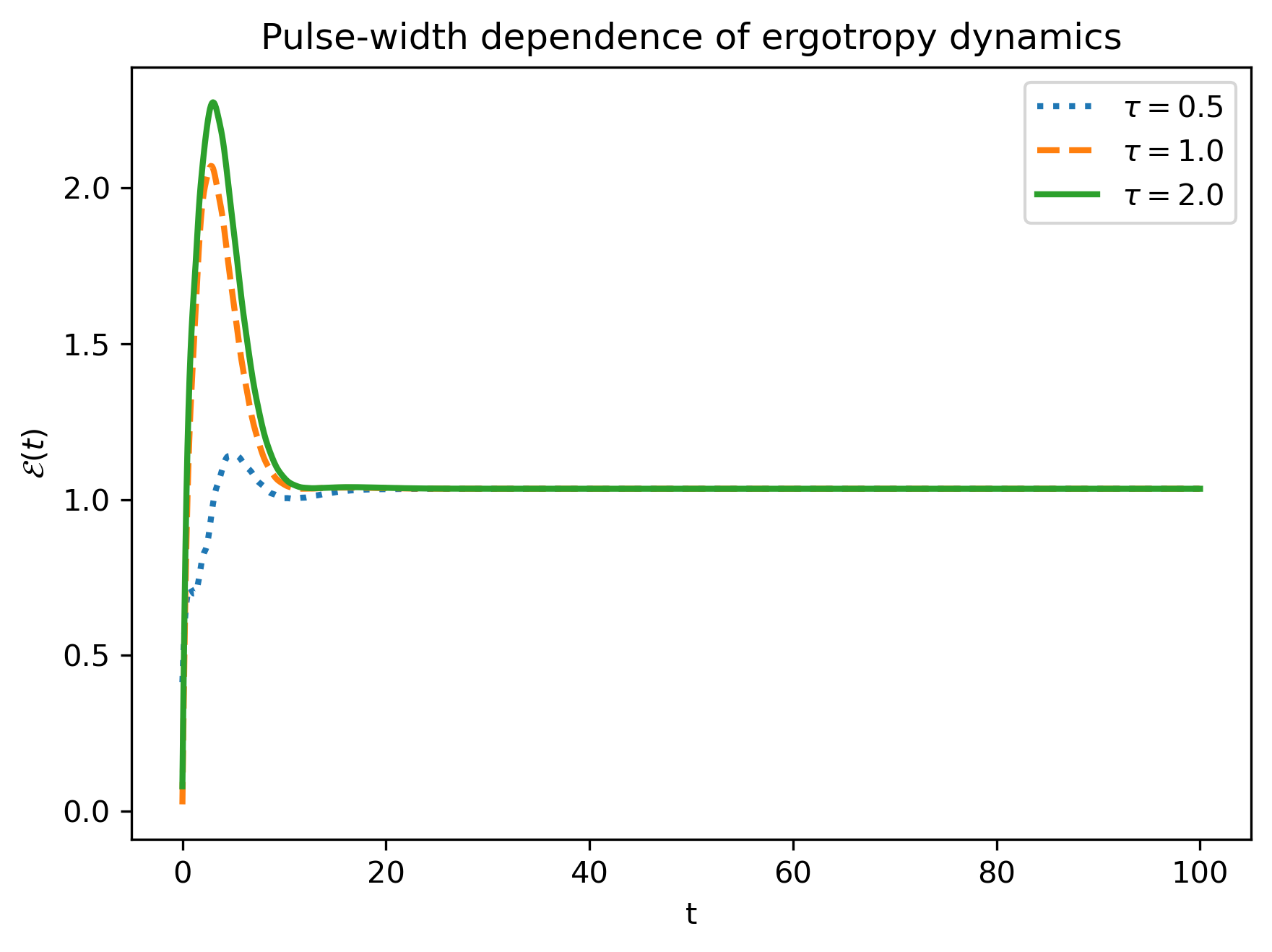}
   \caption{Time evolution of ergotropy $\mathcal{E}(t)$ for different pulse widths $\tau$ under amplitude damping with $\gamma=0.5$. }
    \label{fig:ref}
\end{figure}

To assess the robustness of the charging protocol, we analyze the ergotropy
dynamics for different pulse widths $\tau$ in the presence of amplitude
damping ($\gamma=0.5$). In this case, the charging pulse and dissipative
evolution act simultaneously, allowing us to directly examine how the pulse
duration influences the transient charging dynamics and peak ergotropy.
As shown in Fig.~\ref{fig:ref}, shorter pulses lead to faster but less selective excitation, resulting in reduced peak ergotropy, whereas broader pulses enable smoother and more coherent energy transfer, yielding higher transient ergotropy.

In the long-time regime, however, the ergotropy approaches nearly the same steady-state value for all $\tau$, indicating that the pulse width primarily influences the transient charging behavior rather than the asymptotic dissipative state.
The sensitivity to intrinsic system parameters, such as the coupling strength can be understood through their influence on the level spacing and hybridization of the four-level system. Variations in these parameters modify the efficiency of population transfer and peak ergotropy, but do not qualitatively alter the dissipative dynamics discussed here.
\subsection{ Amplitude Damping and Dephasing}
\label{subsec:addeph}
In the dissipative phase following coherent charging, two types of environmental noise are considered: amplitude damping (AD) and pure dephasing (Deph). These noise channels are modeled using Lindblad-type master equations with distinct collapse operators that characterize their physical effects.
For amplitude damping, the system loses both energy and coherence due to irreversible excitation decay. The corresponding collapse operator is given by eq.(\ref{eq:ad}), where \(\sigma_- \otimes I\) denotes amplitude damping on the first qubit while \(I \otimes \sigma_- \) applies it to the second. Their sum models collective amplitude damping on both qubits. This operator captures the decay processes that remove excitations from the system~\cite{gardiner2004quantum, breuer2002theory}.

In contrast, dephasing is modeled using the collapse operator 
\begin{equation}
 L_{\text{Deph}} = (\sigma_z \otimes I + I \otimes \sigma_z) 
\end{equation} which affects only the coherence (off-diagonal elements of the density matrix) while preserving the populations. Since the operator is diagonal in the computational basis \(\ket{00}, \ket{01}, \ket{10}, \ket{11}\), it results in pure phase decoherence without energy loss~\cite{nielsen2010quantum, schlosshauer2007decoherence}.

\begin{figure}[H]
    \centering
    \includegraphics[width=0.9\linewidth]{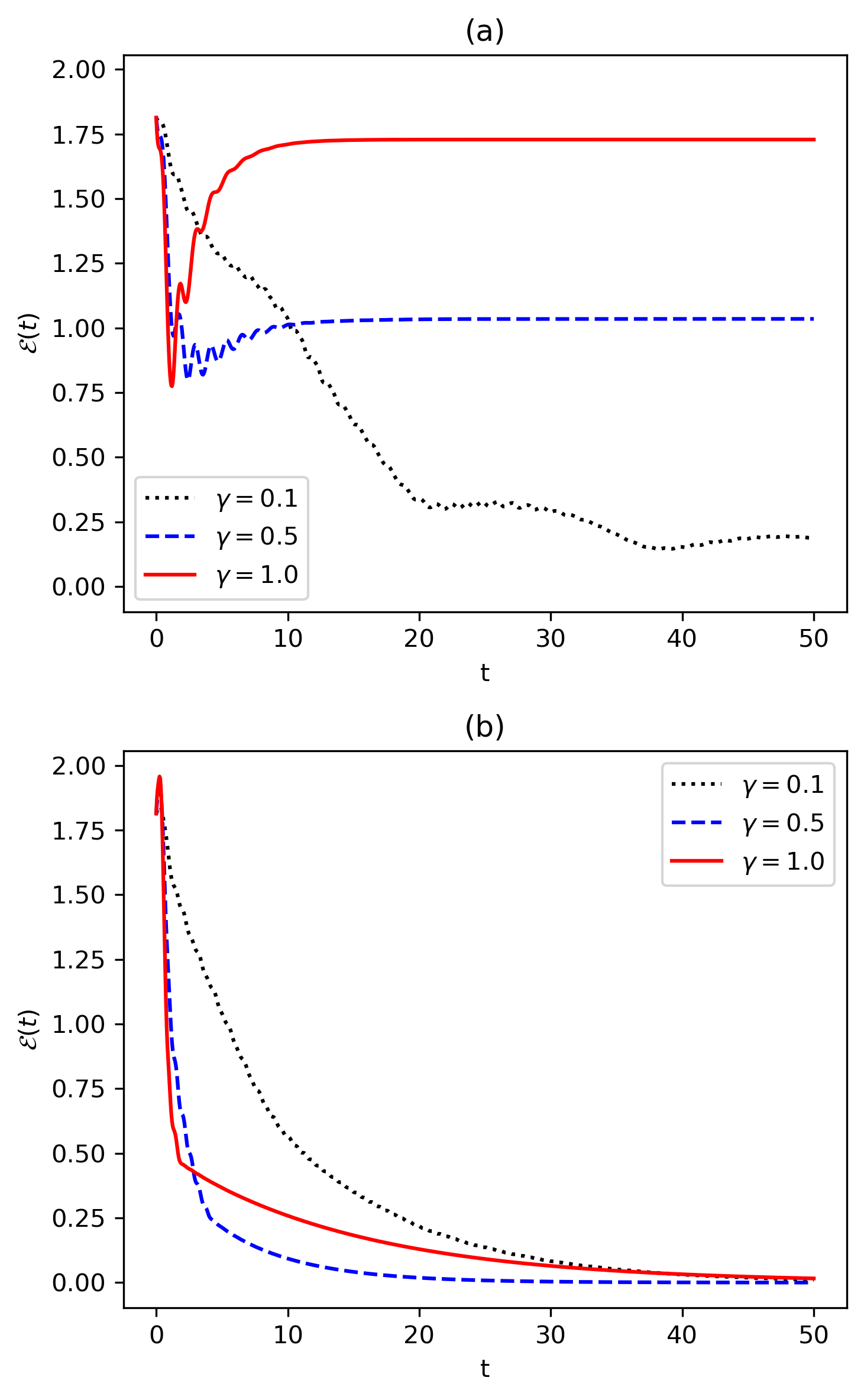}
    \caption{Comparison of ergotropy evolution for different dissipation rates $\gamma$ under (a) amplitude damping and (b) dephasing noise models.}
    \label{fig:comparison}
\end{figure}

Fig.\ref{fig:comparison} presents a comparative analysis of the time evolution of ergotropy for a graphene-based quantum battery subject to amplitude damping (AD) and pure dephasing (Deph) at representative dissipation strengths $\gamma = 0.1$, $0.5$, and $1.0$. In the AD case Fig.\ref{fig:comparison}(a), all $\gamma$ values exhibit an initial ergotropy build-up due to coherent charging, followed by decay as dissipation acts. For weak damping ($\gamma=0.1$), excitations leak slowly into the reservoir and coherence decays gradually, driving the state toward passivity with negligible long-time ergotropy ($\mathcal{E}(t)  \approx 0.18$). At intermediate damping ($\gamma=0.5$), destructive coherences are suppressed while population imbalance persists, resulting in a finite ergotropy plateau. Remarkably, strong damping ($\gamma=1.0$) yields the \emph{highest} steady-state ergotropy, as rapid dissipation freezes favorable population distributions despite substantial coherence loss.  

In contrast, Deph Fig.~\ref{fig:comparison}(b) consistently destroys quantum advantage: coherence collapses rapidly ($C_{\ell_1} \to 0$) and the system diagonalizes in the energy eigenbasis, evolving into a maximally mixed, fully passive state with vanishing ergotropy, irrespective of $\gamma$. This leads to a strong decay in ergotropy to go from $\mathcal{E}(t) \approx 1.81$ at $t=0$ to $\mathcal{E}(t) \approx 0.0001$ at $t=100$ for a weak dissipation ($\gamma=0.1$) and $\mathcal{E}(t) \approx 0.0008$ at $t=100$ for a strong dissipation ($\gamma=1.0$).

These contrasting behaviors underscore the thermodynamic distinction between coherence-destroying and energy-relaxing channels. While amplitude damping involves energy leakage, it can paradoxically \emph{enhances} ergotropy retention by stabilizing structured, non-passive steady states. Dephasing, despite conserving energy, irreversibly eliminates coherence and forces equilibration into passive configurations with negligible extractable work. Thus, ergotropy preservation is governed not solely by weak system–environment coupling but crucially by the \emph{nature of dissipation}: AD can be harnessed as a resource for long-time work storage, whereas Deph invariably suppresses it~\cite{allahverdyan2004maximal, choquehuanca2024qubit, friis2018precision}.

Thus the effects of disorder and thermal fluctuations are incorporated at a phenomenological level through dephasing and amplitude-damping channels. While these processes generally suppress coherence and reduce ergotropy, the dissipation-assisted stabilization of non-passive states remains qualitatively robust under moderate perturbations.
\subsection{Markovian vs Non-Markovian Dissipation}
\label{subsec:nonmarkov}

To understand how environmental memory influences the performance of the graphene quantum battery, we compare the dynamics generated by a standard Markovian Lindblad master equation with those arising from a time-dependent, non-Markovian dissipation model. The Markovian evolution corresponds to a constant decay rate $\gamma$ and describes a memoryless reservoir that irreversibly drains excitations and coherence. In contrast, the non-Markovian scenario incorporates temporal correlations through a decay rate of the form
\begin{equation}
    \gamma(t) = \gamma_0 e^{-\beta t} \cos(\omega t),
\end{equation}
which enables information and energy to intermittently flow back into the system. The parameters $\gamma_0$, $\beta$, and $\omega$ respectively control the initial coupling strength, the decay of memory, and the oscillatory character of the reservoir correlations. Each value of $\beta$ therefore corresponds to a distinct memory timescale, with smaller $\beta$ indicating strong and long-lasting environmental memory.

\subsubsection{Energy Dynamics.}
\begin{figure}[H]
    \centering
    \includegraphics[width=0.9\linewidth]{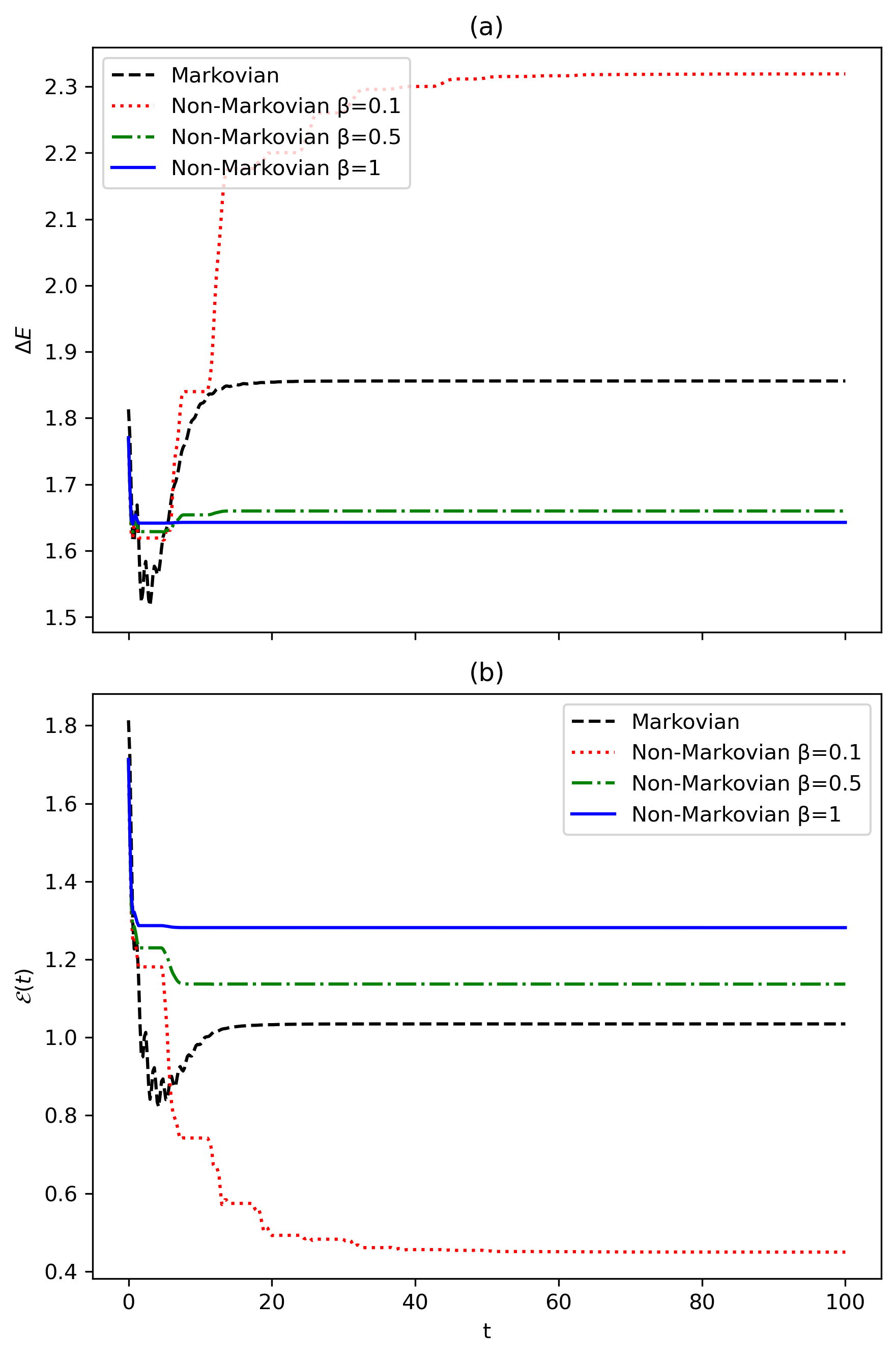}
    \caption{Comparison of energy and ergotropy evolution for markovian and non markovian (solid for $\beta$ = 0.1, 0.5, 1.0 ) environment ($\gamma_0=0.5$, $\omega=1$).}
    \label{fig:memory}
\end{figure}

Figure~\ref{fig:memory}(a) shows the evolution of the stored energy under Markovian and  non-Markovian dissipation. In the Markovian case (black dashed),  the stored energy exhibits a brief transient decrease at short times,  followed by recovery and stabilization at an intermediate steady value  $\Delta E \approx 1.85$. This behavior reflects rapid initial energy  exchange with a memoryless bath and subsequent equilibration governed by a constant decay rate.

The non-Markovian dynamics display a pronounced dependence on the  memory parameter $\beta$. For strong memory ($\beta=0.1$, dotted red),  the stored energy increases significantly after the initial transient and stabilizes at the highest asymptotic value $\Delta E \approx 2.3$. This enhancement indicates substantial  information and energy backflow from the environment, effectively reinforcing the battery’s stored energy. 

For moderate memory strengths ($\beta=0.5$ and $1.0$), the decay rate varies more rapidly in time, reducing the duration of system–bath correlations. Consequently, the steady-state stored energy settles at 
lower values ($\Delta E \approx 1.65$–$1.66$), slightly below the Markovian case, and exhibits smoother relaxation without large backflow-induced amplification. 

These results demonstrate that the stored energy is highly sensitive to environmental memory effects: strong memory enhances long-time energy retention through backflow, whereas shorter memory times lead 
to more conventional relaxation toward lower steady-state values.

\subsubsection{Ergotropy Dynamics.}
Figure~\ref{fig:memory}(b) shows the ergotropy under Markovian and non-Markovian dissipation. In the Markovian case (black dashed), ergotropy rapidly decreases to values around $0.8$ before gradually 
recovering and stabilizing near $1.05$, reflecting irreversible coherence loss in a memoryless environment. The non-Markovian profiles exhibit a different behavior. For strong memory 
($\beta = 0.1$), the system remains strongly coupled to the environment for an extended duration, and the initial backflow is insufficient to preserve useful work; consequently, the ergotropy steadily decays and 
saturates near $0.4$. For moderate memory strengths ($\beta = 0.5$ and $1.0$), however, the time-dependent decay rate retains significant value while still allowing for coherence revival. In these 
regimes, information backflow counteracts dissipation, enabling the ergotropy to stabilize at values ($1.15$--$1.30$) substantially higher than those achieved under Markovian evolution. These cases represent  non-Markovian enhancement, as the environment's finite memory actively contributes to preserving coherence and supporting extractable work.

The observed behavior in ergotropy serves as an operational signature of non-Markovian dynamics, reflecting information backflow and partial preservation of coherence (Appendix C). In particular, moderate non-Markovian memory enhances long-time ergotropy retention compared to the Markovian case, demonstrating the beneficial role of environmental correlations in preserving extractable work.

In summary, For comparison between Markovian and non-Markovian dynamics, we analyze the maximum ergotropy and long-time retention as measures of extractable work. The results show that non-Markovian memory effects enhance work preservation and improve long-time quantum battery performance.

The plotted results reveal a nontrivial interplay between environmental memory and quantum battery performance. While extremely strong memory ($\beta = 0.1$) enhances energy inflow but suppresses long-time extractable work, moderate non-Markovian memory ($\beta = 0.5, 1.0$) enables higher retained ergotropy than the Markovian case.

\section{Conclusion}

In this work, we have analyzed the dissipative dynamics of a pulse-driven graphene-based quantum battery modeled as a four-level spin–valley coupled system. Charging was implemented via a Gaussian pulse, which enabled coherent population transfer and efficient energy storage under closed-system dynamics. The study of open-system effects revealed the intricate interplay between energy dissipation, coherence degradation, and ergotropy retention. 

Our analysis of different noise channels shows that amplitude damping (AD) and dephasing (DP) have qualitatively distinct impacts. While AD causes both energy leakage and coherence decay, it can stabilize non-passive steady states that retain finite ergotropy through population asymmetry. In contrast, DP conserves energy but suppresses coherence entirely, leading to rapid ergotropy loss even when finite internal energy is maintained. This highlights the central role of coherence, rather than energy alone, in governing useful work extraction from quantum batteries.

A comparison of Markovian and non-Markovian dynamics highlights the critical role of reservoir structure in determining long-time battery performance. In the Markovian regime, both energy and ergotropy exhibit 
rapid initial decay followed by relaxation to modest steady-state values, a signature of irreversible dissipation into a memoryless bath. Introducing non-Markovianity through a time-dependent decay rate 
substantially alters this behavior. Non-Markovianity enhances long-time ergotropy, when the decay rate retains memory long enough to permit information backflow but still decays sufficiently fast to suppress prolonged dissipation. Very small $\beta$ leads to strong dissipation, and very large $\beta$ effectively increases the ergotropy of the system.

The present four-level Hamiltonian provides a minimal effective description of a finite graphene subsystem, capturing the essential interplay between coherent driving, dissipation, and ergotropy. While this framework successfully describes the fundamental mechanisms of energy storage and dissipation-assisted stabilization at the single-unit level, it does not explicitly incorporate effects such as spatial inhomogeneity, strong disorder, or many-body correlations that may arise in larger graphene structures or coupled multi-cell configurations. These additional degrees of freedom can introduce cooperative charging effects and correlated relaxation processes, potentially modifying both the transient dynamics and long-time ergotropy retention. 

Taken together, these findings establish that dissipation is not merely a destructive factor but can be engineered as a resource. Amplitude damping can induce ergotropy stabilization, and non-Markovian memory effects can prolong useful work extraction. Such insights provide a concrete pathway toward reservoir-engineered graphene quantum batteries, where the combination of coherent charging protocols and structured environments can yield robust nanoscale energy storage devices with enhanced operational efficiency. A possible extension of the present work would involve incorporating many-body interactions and coupled multi-cell graphene quantum battery networks to investigate collective charging effects and correlated dissipative dynamics.

\appendix
\section*{Appendix A: Density-Matrix Diagnostics}

To validate the main text results in Sec. ~\ref{subsec:erg}, we present tabulated values of the
$\ell_{1}$-coherence $C_{\ell_1}$, ergotropy $\mathcal{E}(t)$, and eigenvalue
distributions of $\rho(t)$ for selected times $t$ and dissipation
strengths $\gamma$. These data are obtained directly from numerical
simulations and illustrate how population redistribution and coherence
loss govern ergotropy retention.

\begin{table}[H]
\centering
\caption{Weak dissipation ($\gamma=0.1$): time evolution of coherence,
ergotropy, and eigenvalue spectrum of $\rho(t)$.}
\begin{tabular}{c c c c}
\hline
$t$ & $C_{\ell_1}$ & $\mathcal{E}(t)$ & Eigenvalues of $\rho(t)$ \\
\hline
0   & 2.1013 & 1.8133 & \{0.353, 0.001, 0.355, 0.291\} \\
10  & 1.3676 & 0.9180 & \{0.33, 0.23, 0.182, 0.258\} \\
40  & 0.5145 & 0.1447 & \{0.265, 0.231, 0.282, 0.222\} \\
100 & 0.4705 & 0.1777 & \{0.270, 0.228, 0.271, 0.231\} \\
\hline
\end{tabular}
\end{table}

\begin{table}[H]
\centering
\caption{Intermediate dissipation ($\gamma=0.5$): time evolution of coherence,
ergotropy, and eigenvalue spectrum of $\rho(t)$.}
\begin{tabular}{c c c c}
\hline
$t$ & $C_{\ell_1}$ & $\mathcal{E}(t)$ & Eigenvalues of $\rho(t)$ \\
\hline
0   & 2.1013 & 1.8133 & \{0.353, 0.001, 0.355, 0.291\} \\
10  & 1.0078 & 0.9823 & \{0.407, 0.099, 0.384, 0.11\} \\
40  & 0.9983 & 1.0347 & \{0.401, 0.103, 0.393, 0.103\} \\
100 & 0.9983 & 1.0347 & \{0.401, 0.103, 0.393, 0.103\} \\
\hline
\end{tabular}
\end{table}

\begin{table}[H]
\centering
\caption{Strong dissipation ($\gamma=1.0$): time evolution of coherence,
ergotropy, and eigenvalue spectrum of $\rho(t)$.}
\begin{tabular}{c c c c}
\hline
$t$ & $C_{\ell_1}$ & $\mathcal{E}(t)$ & Eigenvalues of $\rho(t)$ \\
\hline
0   & 2.1013 & 1.8133 & \{0.353, 0.001, 0.355, 0.291\} \\
10  & 0.8960 & 1.7131 & \{0.453, 0.073, 0.416, 0.058\} \\
40  & 0.8923 & 1.7284 & \{0.453, 0.073, 0.416, 0.058\} \\
100 & 0.8923 & 1.7284 & \{0.453, 0.073, 0.416, 0.058\} \\
\hline
\end{tabular}
\end{table}
\FloatBarrier
\appendix
\section*{Appendix B: Dissipation-Channel Diagnostics}

To complement the statement in Sec.\ref{subsec:addeph}, we tabulate the $\ell_{1}$-coherence 
$C_{\ell_1}$, ergotropy $\mathcal{E}(t)$, and eigenvalue distributions of $\rho(t)$ 
for representative times under amplitude damping (AD) and dephasing 
(Deph) channels. These results demonstrate how the two noise models 
differ in their impact on coherence retention, population redistribution, 
and long-time ergotropy.

\begin{table}[H]

\centering
\caption{Dephasing (Deph) with weak dissipation ($\gamma=0.1$).}
\begin{tabular}{c c c c}
\hline
$t$ & $C_{\ell_1}$ & $\mathcal{E}(t)$ & Eigenvalues of $\rho(t)$ \\
\hline
0   & 2.1013 & 1.8133 & \{0.353, 0.001, 0.355, 0.291\} \\
10  & 0.3469 & 0.5564 & \{0.283, 0.308, 0.186, 0.223\} \\
40  & 0.0160 & 0.0315 & \{0.252, 0.251, 0.249, 0.248\} \\
100 & 0.0000 & 0.0001 & \{0.250, 0.250, 0.250, 0.250\} \\
\hline
\end{tabular}
\end{table}

\begin{table}[H]
\centering
\caption{Dephasing (Deph) with strong dissipation ($\gamma=1.0$).}
\begin{tabular}{c c c c}
\hline
$t$ & $C_{\ell_1}$ & $\mathcal{E}(t)$ & Eigenvalues of $\rho(t)$ \\
\hline
0   & 2.1013 & 1.8133 & \{0.353, 0.001, 0.355, 0.291\} \\
10  & 0.0888 & 0.4248 & \{0.288, 0.222, 0.278, 0.212\} \\
40  & 0.0111 & 0.0529 & \{0.255, 0.246, 0.254, 0.245\} \\
100 & 0.0002 & 0.0008 & \{0.25, 0.25, 0.25, 0.25\} \\
\hline
\end{tabular}
\end{table}

\appendix
\section*{Appendix C: Markovian and Non - markovian Channel  Diagnostics}
To complement the analysis in Sec.~\ref{subsec:nonmarkov}, we present 
representative values of the $\ell_{1}$-coherence $C_{\ell_1}$, ergotropy 
$\mathcal{E}(t)$, and eigenvalue spectra of $\rho(t)$ for Markovian and 
Non-Markovian reservoirs with $\beta = 0.1, 0.5, 1.0, 5.0$. These results 
highlight how memory strength influences coherence retention and 
population reshaping, thereby determining the long-time work-storage 
performance.

\begin{table}[H]
\centering
\caption{(a) Markovian}
\begin{tabular}{c c c c}
\hline
$t$ & $C_{\ell_1}$ & $\mathcal{E}(t)$ & Eigenvalues of $\rho(t)$ \\
\hline
0   & 2.1012 & 1.8133 & \{0.353, 0.001, 0.355, 0.291\} \\
10  & 1.0066 & 0.9824 & \{0.407, 0.099, 0.384, 0.109\} \\
40  & 0.9983 & 1.0347 & \{0.401, 0.103, 0.393, 0.103\} \\
100 & 0.9983 & 1.0347 & \{0.401, 0.103, 0.393, 0.103\} \\
\hline
\end{tabular}
\end{table}

\begin{table}[H]
\centering
\caption{(b) Non-Markovian ($\beta=0.1$)}
\begin{tabular}{c c c c}
\hline
$t$ & $C_{\ell_1}$ & $\mathcal{E}(t)$ &  Eigenvalues of $\rho(t)$ \\
\hline
0   & 2.5854 & 1.7132 & \{0.226, 0.115, 0.376, 0.282\} \\
10  & 1.3513 & 0.7423 & \{0.242, 0.240, 0.232, 0.286\} \\
40  & 0.6760 & 0.4560 & \{0.211, 0.236, 0.324, 0.228\} \\
100 & 0.8107 & 0.4496 & \{0.245, 0.255, 0.239, 0.262\} \\
\hline
\end{tabular}
\end{table}

\begin{table}[H]
\centering
\caption{(c) Non-Markovian ($\beta=0.5$)}
\begin{tabular}{c c c c}
\hline
$t$ & $C_{\ell_1}$ & $\mathcal{E}(t)$ &  Eigenvalues of $\rho(t)$ \\
\hline
0   & 2.5854 & 1.7132 & \{0.226, 0.115, 0.376, 0.282\} \\
10  & 2.1277 & 1.1371 & \{0.213, 0.284, 0.186, 0.317\} \\
40  & 1.7619 & 1.1369 & \{0.075, 0.232, 0.429, 0.265\} \\
100 & 2.1614 & 1.1368 & \{0.264, 0.226, 0.266, 0.244\} \\
\hline
\end{tabular}
\end{table}

\begin{table}[H]
\centering
\caption{(d) Non-Markovian ($\beta=1.0$)}
\begin{tabular}{c c c c}
\hline
$t$ & $C_{\ell_1}$ & $\mathcal{E}(t)$ &  Eigenvalues of $\rho(t)$ \\
\hline
0   & 2.5854 & 1.7132 & \{0.226, 0.115, 0.376, 0.282\} \\
10  & 2.3058 & 1.2817 & \{0.219, 0.278, 0.191, 0.311\} \\
40  & 1.8715 & 1.2817 & \{0.050, 0.230, 0.447, 0.272\} \\
100 & 2.3346 & 1.2816 & \{0.264, 0.219, 0.276, 0.241\} \\
\hline
\end{tabular}
\end{table}



\begin{thebibliography}{10}

\bibitem{lu2024peptide}
Jiao~Yang Lu, Zhen Guo, Wei~Tao Huang, Meihua Bao, Binsheng He, Guangyi Li, Jieni Lei, and Yaqian Li.
\newblock Peptide-graphene logic sensing system for dual-mode detection of exosomes, molecular information processing and protection.
\newblock {\em Talanta}, 267:125261, 2024.

\bibitem{peng2025ultrahigh}
Zhuo-Xun Peng, Bo-Xun Li, and Chao-Sheng Deng.
\newblock Ultrahigh-q fano resonance in a cavity-waveguide coupled system based on second-order topological photonic crystals with elliptical holes.
\newblock {\em Optics \& Laser Technology}, 181:111617, 2025.

\bibitem{zhang2025dominant}
Yunfei Zhang, Jian Li, Mingzhe Yu, Xu~Chen, Xingying Chen, and Jun Shen.
\newblock Dominant factor identification and fast optimization of carnot battery by integrating shap and physics-guided neural network.
\newblock {\em Applied Energy}, 401:126641, 2025.

\bibitem{novoselov2004electric}
Kostya~S Novoselov, Andre~K Geim, Sergei~V Morozov, De-eng Jiang, Yanshui Zhang, Sergey~V Dubonos, Irina~V Grigorieva, and Alexandr~A Firsov.
\newblock Electric field effect in atomically thin carbon films.
\newblock {\em science}, 306(5696):666--669, 2004.

\bibitem{semenoff1984condensed}
Gordon~W Semenoff.
\newblock Condensed-matter simulation of a three-dimensional anomaly.
\newblock {\em Physical Review Letters}, 53(26):2449, 1984.

\bibitem{divincenzo1984self}
DP~DiVincenzo and EJ~Mele.
\newblock Self-consistent effective-mass theory for intralayer screening in graphite intercalation compounds.
\newblock {\em Physical Review B}, 29(4):1685, 1984.

\bibitem{fradkin1986critical}
Eduardo Fradkin.
\newblock Critical behavior of disordered degenerate semiconductors. ii. spectrum and transport properties in mean-field theory.
\newblock {\em Physical review B}, 33(5):3263, 1986.

\bibitem{haldane1988model}
F~Duncan~M Haldane.
\newblock Model for a quantum hall effect without landau levels: Condensed-matter realization of the" parity anomaly".
\newblock {\em Physical review letters}, 61(18):2015, 1988.

\bibitem{novoselov2005two}
Kostya~S Novoselov, Andre~K Geim, Sergei~Vladimirovich Morozov, Dingde Jiang, Michail~I Katsnelson, Irina~V Grigorieva, Sergey~V Dubonos, and Alexandr~A Firsov.
\newblock Two-dimensional gas of massless dirac fermions in graphene.
\newblock {\em nature}, 438(7065):197--200, 2005.

\bibitem{verma2025dynamics}
Disha Verma, VS~Indrajith, and R~Sankaranarayanan.
\newblock Dynamics of heisenberg xyz two-spin quantum battery.
\newblock {\em Physica A: Statistical Mechanics and its Applications}, 659:130352, 2025.

\bibitem{le2018spin}
Thao~P Le, Jesper Levinsen, Kavan Modi, Meera~M Parish, and Felix~A Pollock.
\newblock Spin-chain model of a many-body quantum battery.
\newblock {\em Physical Review A}, 97(2):022106, 2018.

\bibitem{dou2022cavity}
Fu-Quan Dou, Hang Zhou, and Jian-An Sun.
\newblock Cavity heisenberg-spin-chain quantum battery.
\newblock {\em Physical Review A}, 106(3):032212, 2022.

\bibitem{gemme2023off}
Giulia Gemme, Gian~Marcello Andolina, Francesco Maria~Dimitri Pellegrino, Maura Sassetti, and Dario Ferraro.
\newblock Off-resonant dicke quantum battery: Charging by virtual photons.
\newblock {\em Batteries}, 9(4):197, 2023.

\bibitem{catalano2024frustrating}
Alberto~Giuseppe Catalano, Salvatore~Marco Giampaolo, Oliver Morsch, Vittorio Giovannetti, and Fabio Franchini.
\newblock Frustrating quantum batteries.
\newblock {\em PRX Quantum}, 5(3):030319, 2024.

\bibitem{cavaliere2025gaussian}
Fabio Cavaliere, Dario Ferraro, Matteo Carrega, Giuliano Benenti, and Maura Sassetti.
\newblock Quantum advantage bounds for a multipartite gaussian battery.
\newblock {\em arXiv preprint arXiv:2510.24162}, 2025.

\bibitem{grazi2025fermion}
Riccardo Grazi, Fabio Cavaliere, Maura Sassetti, Dario Ferraro, and Niccol{\`o} Traverso~Ziani.
\newblock Charging free fermion quantum batteries.
\newblock {\em Chaos, Solitons \& Fractals}, 196:116383, 2025.

\bibitem{massa2025collisional}
Nicol{\`o} Massa, Fabio Cavaliere, and Dario Ferraro.
\newblock The collisional charging of a transmon quantum battery.
\newblock {\em Batteries}, 11(7):240, 2025.

\bibitem{cavaliere2025blockade}
Fabio Cavaliere, Giacomo Gemme, Giuliano Benenti, Dario Ferraro, and Maura Sassetti.
\newblock Dynamical blockade of a reservoir for optimal performances of a quantum battery.
\newblock {\em Communications Physics}, 8(1):76, 2025.

\bibitem{grazi2026quenches}
Riccardo Grazi, Dario Farina, Niccol{\`o} Traverso~Ziani, and Dario Ferraro.
\newblock Charging quantum batteries via dissipative quenches.
\newblock {\em arXiv preprint arXiv:2604.08151}, 2026.

\bibitem{chand2026spin}
S.~Chand, R.~Grazi, Niccol{\`o} Traverso~Ziani, and Dario Ferraro.
\newblock Spin-based quantum energy devices: From quantum thermal machines to quantum batteries.
\newblock {\em Entropy}, 28(4):396, 2026.

\bibitem{farina2026phase}
Dario Farina, Maura Sassetti, Vittorio Cataudella, Dario Ferraro, and Niccol{\`o} Traverso~Ziani.
\newblock Charging power enhancement at the phase transition of a non-integrable quantum battery.
\newblock {\em arXiv preprint arXiv:2603.02819}, 2026.

\bibitem{pavone2026cluster}
Alessandro Pavone, Francesco~Leonardo Cavagnaro, Matteo Carrega, Riccardo Grazi, Dario Ferraro, and Niccol{\`o} Traverso~Ziani.
\newblock Cluster ising quantum batteries can mimic super-extensive charging power.
\newblock {\em arXiv preprint arXiv:2602.15467}, 2026.

\bibitem{grazi2025finite}
Riccardo Grazi, Henrik Johannesson, Dario Ferraro, and Niccol{\`o} Traverso~Ziani.
\newblock Finite-time protocols stabilize charging in noisy ising quantum batteries.
\newblock {\em arXiv preprint arXiv:2512.14521}, 2025.

\bibitem{HU2025116229}
Weiran Hu, Shuochen Yang, Jiangfeng Tian, Zirong He, Liang Qiu, and Fangxin Zhang.
\newblock The performance of quantum battery in a common dephasing environment.
\newblock {\em Physica E: Low-dimensional Systems and Nanostructures}, 170:116229, 2025.

\bibitem{kamin2020nonmarkovian}
Felix~H. Kamin, Seyed Salimi, Jader~P. Santos, Gabriel~T. Landi, and Kamil Korzekwa.
\newblock Non-markovian effects on the charging and self-discharging process of quantum batteries.
\newblock {\em New Journal of Physics}, 22(8):083007, 2020.

\bibitem{zhao2021environmental}
Fang Zhao, Fulai Dou, Quan Shi, and Xiaotao Wu.
\newblock Quantum battery of interacting spins with environmental noise.
\newblock {\em Physical Review A}, 104(3):033715, 2021.

\bibitem{zakavati2021bounds}
Sajad Zakavati, Antonio~M. Timpanaro, and Gabriel~T. Landi.
\newblock Bounds on charging power of open quantum batteries.
\newblock {\em Physical Review E}, 104(3):034117, 2021.

\bibitem{quach2020dark}
James~Q. Quach and William~J. Munro.
\newblock Using dark states to charge and stabilize open quantum batteries.
\newblock {\em Physical Review Applied}, 14(2):024092, 2020.

\bibitem{quach2020using}
James~Q Quach and William~J Munro.
\newblock Using dark states to charge and stabilize open quantum batteries.
\newblock {\em Physical Review Applied}, 14(2):024092, 2020.

\bibitem{liu2019loss}
Junjie Liu, Dvira Segal, and Gabriel Hanna.
\newblock Loss-free excitonic quantum battery.
\newblock {\em The Journal of Physical Chemistry C}, 123(30):18303--18314, 2019.

\bibitem{campaioli2024colloquium}
Francesco Campaioli, Stefano Gherardini, James~Q Quach, Marco Polini, and Gian~Marcello Andolina.
\newblock Colloquium: quantum batteries.
\newblock {\em Reviews of Modern Physics}, 96(3):031001, 2024.

\bibitem{eich2018spin}
Marius Eich, Franti{\v{s}}ek Herman, Riccardo Pisoni, Hiske Overweg, Annika Kurzmann, Yongjin Lee, Peter Rickhaus, Kenji Watanabe, Takashi Taniguchi, Manfred Sigrist, et~al.
\newblock Spin and valley states in gate-defined bilayer graphene quantum dots.
\newblock {\em Physical Review X}, 8(3):031023, 2018.

\bibitem{yang2013spin}
CH~Yang, A~Rossi, R~Ruskov, NS~Lai, FA~Mohiyaddin, S~Lee, C~Tahan, Gerhard Klimeck, A~Morello, and AS~Dzurak.
\newblock Spin-valley lifetimes in a silicon quantum dot with tunable valley splitting.
\newblock {\em Nature communications}, 4(1):2069, 2013.

\bibitem{beenakker2006specular}
CWJ Beenakker.
\newblock Specular andreev reflection in graphene.
\newblock {\em Physical review letters}, 97(6):067007, 2006.

\bibitem{traversoziani2015josephson}
Niccol\`o Traverso~Ziani et~al.
\newblock Josephson current in ballistic graphene sns junctions.
\newblock {\em Physical Review B}, 92:245428, 2015.

\bibitem{cavaliere2018andreev}
Fabio Cavaliere et~al.
\newblock Andreev bound states and coherent transport in hybrid superconducting nanostructures.
\newblock {\em Physica E}, 99:192--198, 2018.

\bibitem{mak2014valley}
Kin~Fai Mak, Kathryn~L McGill, Jiwoong Park, and Paul~L McEuen.
\newblock The valley hall effect in mos2 transistors.
\newblock {\em Science}, 344(6191):1489--1492, 2014.

\bibitem{binder2018thermodynamics}
Felix Binder, Luis~A Correa, Christian Gogolin, Janet Anders, and Gerardo Adesso.
\newblock Thermodynamics in the quantum regime.
\newblock {\em Fundamental Theories of Physics}, 195(1), 2018.

\bibitem{paulson2025work}
Kavalambramalil~George Paulson, Hanna Terletska, and Herbert~F Fotso.
\newblock Work extraction from a controlled quantum emitter.
\newblock {\em Journal of Physics: Photonics}, 7(2):025023, 2025.

\bibitem{barra2019dissipative}
Felipe Barra.
\newblock Dissipative charging of a quantum battery.
\newblock {\em Physical review letters}, 122(21):210601, 2019.

\bibitem{hu2009quantum}
Zhan-Ning Hu, Kee-Su Park, and Kyung-Soo Yi.
\newblock Quantum entanglement in a graphene sheet.
\newblock {\em J. Korean Phys. Soci}, 54:921, 2009.

\bibitem{quach2022superabsorption}
James~Q Quach, Kirsty~E McGhee, Lucia Ganzer, Dominic~M Rouse, Brendon~W Lovett, Erik~M Gauger, Jonathan Keeling, Giulio Cerullo, David~G Lidzey, and Tersilla Virgili.
\newblock Superabsorption in an organic microcavity: Toward a quantum battery.
\newblock {\em Science advances}, 8(2):eabk3160, 2022.

\bibitem{ferraro2018high}
Dario Ferraro, Michele Campisi, Gian~Marcello Andolina, Vittorio Pellegrini, and Marco Polini.
\newblock High-power collective charging of a solid-state quantum battery.
\newblock {\em Physical review letters}, 120(11):117702, 2018.

\bibitem{campaioli2017enhancing}
Francesco Campaioli, Felix~A Pollock, Felix~C Binder, Lucas C{\'e}leri, John Goold, Sai Vinjanampathy, and Kavan Modi.
\newblock Enhancing the charging power of quantum batteries.
\newblock {\em Physical review letters}, 118(15):150601, 2017.

\bibitem{indrajith2022fidelity}
VS~Indrajith, R~Muthuganesan, and R~Sankaranarayanan.
\newblock Fidelity-based purity and coherence for quantum states.
\newblock {\em International Journal of Quantum Information}, 20(06):2250016, 2022.

\bibitem{allahverdyan2004maximal}
Armen~E Allahverdyan, Roger Balian, and Th~M Nieuwenhuizen.
\newblock Maximal work extraction from finite quantum systems.
\newblock {\em Europhysics Letters}, 67(4):565, 2004.

\bibitem{francica2017daemonic}
Gianluca Francica, John Goold, Francesco Plastina, and Mauro Paternostro.
\newblock Daemonic ergotropy: Enhanced work extraction from quantum correlations.
\newblock {\em npj Quantum Information}, 3(1):12, 2017.

\bibitem{breuer2002theory}
Heinz-Peter Breuer and Francesco Petruccione.
\newblock {\em The theory of open quantum systems}.
\newblock OUP Oxford, 2002.

\bibitem{manzano2020short}
Daniel Manzano.
\newblock A short introduction to the lindblad master equation.
\newblock {\em Aip advances}, 10(2), 2020.

\bibitem{nielsen2010quantum}
Michael~A Nielsen and Isaac~L Chuang.
\newblock {\em Quantum computation and quantum information}.
\newblock Cambridge university press, 2010.

\bibitem{baumgratz2014quantifying}
Tillmann Baumgratz, Marcus Cramer, and Martin~B Plenio.
\newblock Quantifying coherence.
\newblock {\em Physical review letters}, 113(14):140401, 2014.

\bibitem{ccakmak2020ergotropy}
Bar{\i}{\c{s}} {\c{C}}akmak.
\newblock Ergotropy from coherences in an open quantum system.
\newblock {\em Physical Review E}, 102(4):042111, 2020.

\bibitem{ghosh2021fast}
Srijon Ghosh, Titas Chanda, Shiladitya Mal, and Aditi Sen.
\newblock Fast charging of a quantum battery assisted by noise.
\newblock {\em Physical Review A}, 104(3):032207, 2021.

\bibitem{gardiner2004quantum}
Crispin Gardiner and Peter Zoller.
\newblock {\em Quantum noise: a handbook of Markovian and non-Markovian quantum stochastic methods with applications to quantum optics}.
\newblock Springer Science \& Business Media, 2004.

\bibitem{schlosshauer2007decoherence}
Maximilian Schlosshauer.
\newblock {\em Decoherence and the quantum-to-classical transition}.
\newblock Springer, 2007.

\bibitem{choquehuanca2024qubit}
JMZ Choquehuanca, PAC Obando, FM~de~Paula, and MS~Sarandy.
\newblock Qubit dynamics of ergotropy and environment-induced work.
\newblock {\em Physical Review A}, 109(5):052219, 2024.

\bibitem{friis2018precision}
Nicolai Friis and Marcus Huber.
\newblock Precision and work fluctuations in gaussian battery charging.
\newblock {\em Quantum}, 2:61, 2018.

\end{thebibliography}
\end{document}